\documentclass[epj,final]{svjour}

\usepackage{bm}

\begin{document}

\title{Spontaneous localization for a charged particle \\
in non-relativistic quantum mechanics}

\author{F. Miglietta \thanks{\email{francesco.miglietta@pv.infn.it}}}

\institute{Dipartimento di Fisica Nucleare e Teorica dell' Universit\`a
di Pavia - via Bassi 6, I - 27100 Pavia - Italy 
\\ I.N.F.N. - Sezione di Pavia}

\date{May 27, 2003 }

\abstract{The interaction of a moving charged particle with its coherent
electromagnetic field is analysed in the framework of non-relativistic quantum
mechanics. It is shown that, when this interaction is taken into account, a
spatially localized state may have a mean energy lower than the one
corresponding to a delocalized state.}


\maketitle

\section{Introduction}

\label{sec:1}

We know, from the study of the infra-red divergence problem in relativistic
QED, that a physical system cannot be separated from its {\it classical} ({\it
i.e.} coherent)  electromagnetic field \cite{1} \cite{2}. On the other hand it
is a well-known fact, in classical physics, that the motion of a charged
particle cannot be described correctly, unless the radiation-reaction force is
taken into account properly \cite{3} . On the contrary, in our opinion, in
non-relativistic quantum mechanics the effects due to the interaction of a
system with the self-generated coherent field have not been analysed in a
satisfactory way.  \par

The aim of this paper consists of an analysis of the  effects due to the
interaction of a moving charged particle with its coherent field.
Uncoherent-photon emission will not be considered here. Radiation damping for a
non-relativistic quantum system has been studied {\it e.g.} in Ref. \cite{4}.
Decoherence effects due to  soft-bremsstrahlung emission have been analysed by
Breuer and Petruccione in the interesting papers of Ref.s \cite{5} and
\cite{6}. \par
 
We will show that, for a moving particle, the interaction with the coherent
field gives origin to a mechanism which favours, from the energetical point of
view, a spontaneous localization in space. The effect is due essentially to the
attractive force which acts among parallel currents. The Coulomb gauge is
assumed in this paper. It is well known that, in the Coulomb gauge,  the
Coulomb interaction is described as an instantaneous action at a distance among
different charged particles. The repulsive Coulomb self-interaction energy
concerning a single charged particle, once the mass-renormalization
contribution (which is independent of the wave function) has been subtracted,
can be estimated of the order of the Lamb-shift energies.  \par

The analysis presented in this paper is based on the fact that, when the
residual interaction with the quantum field is neglected ({\it i.e.} when
uncoherent photon emission is neglected),  a  rather simple expression can be
given for the conserved total energy of the system (particle plus  {\it
classical} field), in terms of the {\it classical} vector potential in the
Coulomb gauge and of the wave function of the particle. We will show that, for
a freely moving electron with {\it e.g.} a convective velocity $ v_c \sim
10^{-1}  c $, a minimum for the energy is attained for a radius of localization
$ b $ of the order of $ 10^{-8} $ m, with a {\it binding} energy $ {\cal E}^b $
of $ 10^{-4} \div 10^{-5} $ eV. For a proton with the same $ v_c $ the result
would be $ b \sim 10^{-11} $ m and $ {\cal E}^b \sim 10^{-1} $ eV. It will be
shown that similar results hold for neutral atoms also.

\section{Classical field and self-interaction}
\label{sec:2}

Let us consider an electron interacting with the transverse electromagnetic
field. For the field we assume the Coulomb gauge, as well as the Schr\"odinger
picture. Let $ \vec{A} $ and $ {\vec{E} }_{\perp } $ be quantum field
operators and let $ {\vec{A} }_c $ and $ {\vec{E} }_{\perp c} $ be the
corresponding {\it classical} fields. The Hamiltonian for the total system,
consisting of the electron and the quantum field, is given by

\begin{equation}
H_T = H_M + H_F + V ,
\end{equation}
where $ H_M $ describes the free electron, $ H_F $ is the free-field
Hamiltonian, given by the following normal-ordered expression 

\begin{equation}
H_F =  \frac{{\epsilon }_0 }{2} \int d \vec{r} \: ( \vert {\vec{E} }_{\perp }
{\vert }^2 + c^2 \vert \vec{B} {\vert }^2 ) :  \:   ,
\end{equation}
and the intraction $ V $ is given by

\begin{equation}
V = - \int d \vec{r} ( \vec{A} \cdot {\vec{j} }_p - \frac{e^2 }{2 m}
\hat{n} \vert \vec{A} {\vert }^2 + \vec{B} \cdot \vec{M}  ) .
\end{equation}
In eq.(3) $ {\vec{j} }_p $ , given by

\begin{equation}
{\vec{j} }_p ( \vec{r} ) = \frac{ i e \hbar }{ 2 m } [ \delta ( \vec{r} -
{\vec{r} }_e ) {\bm{\nabla } }_e + {\bm{\nabla } }_e \delta ( \vec{r} -
{\vec{r} }_e ) ] ,
\end{equation}
represents the {\it canonical} current ({\it i.e.} the current due to the
canonical momentum), $\hat{n}$ represents the electron-density operator and
finally $ \vec{M}  $ represents the spin magnetization-density operator.

The following time-independent commutation relations 

\begin{equation}
[ A_i ( \vec{r} ) , E_{\perp j } ({\vec{r} }' ) ] = - i \hbar {\epsilon
}_0^{-1} {\cal P}_{ij} ( \vec{r} - {\vec{r} }' )
\end{equation}
hold, where

\begin{eqnarray}
&& {\cal P}_{ij} ( \vec{r} ) = ( \frac{1}{2 \pi } )^3 \int d \vec{q} e^{i
\vec{q} \cdot \vec{r} } ( {\delta }_{ij} - \frac{q_i q_j }{q^2 } ) \\
\nonumber 
&&= \frac{2}{3} {\delta }_{ij} \delta ( \vec{r} ) + \frac{3}{4 \pi  r^3} (
\frac{x_i x_j }{r^2 } - \frac{1}{3} {\delta }_{ij} ) .
\end{eqnarray}
is the projector onto the transverse components of the field \cite{7} .
\par

The Schr\"odinger equation describing the total system can be derived from the
variational functional

\begin{equation}
S_Q = \int d t \langle {\Psi }_T \vert {\cal D}^+ ( - i \hbar {\partial }_t +
H_T ) {\cal D} \vert {\Psi }_T \rangle ,
\end{equation}
where the  {\it displacement} unitary operator $ {\cal D} $, given by 

\begin{eqnarray}
{\cal D} ( t ) &&= \exp \{ \frac{i {\epsilon }_0 }{\hbar } \int d \vec{r} [  
{\vec{A} }_c (t,\vec{r} ) \cdot {\vec{E} }_{\perp } ( \vec{r} )
\\ \nonumber 
&&- {\vec{E} }_{\perp c} (t,\vec{r} ) \cdot \vec{A} ( \vec{r} )] \} ,
\end{eqnarray}
has been introduced. The effect of the {\it displacement} operator $ {\cal D} $
on the field operators consists of

\begin{eqnarray}
{\cal D}^{-1}  \vec{A} {\cal D} &&= \vec{A} + {\vec{A} }_c  \\
{\cal D}^{-1} {\vec{E} }_{\perp } {\cal D} &&= {\vec{E} }_{\perp } +
{\vec{E} }_{\perp c} .
\end{eqnarray}

By independent variations of $ S_Q $ with respect to $ {\vec{E} }_{\perp c} 
$ and $ {\vec{A} }_c $ (taking into account the transversality of both) we
obtain the relation

\begin{equation}
{\vec{E} }_{\perp c} = - {\partial }_t{\vec{A} }_c 
\end{equation}
and the Maxwell equation

\begin{equation}
{\epsilon }_0 c^2 \bm{\nabla } \times {\vec{B} }_c - {\epsilon}_0 {\partial
}_t {\vec{E} }_{\perp c } = {\vec{j} }_{c \perp }  ,
\end{equation}
where we have assumed $ \langle \vec{A} \rangle = \vec{0} $ as well as $
\langle {\vec{E} }_{\perp } \rangle = \vec{0} $. The {\it classical} current
$ {\vec{j} }_c $ is given by

\begin{eqnarray}
&& {\vec{j} }_c ( \vec{r} ) = \langle {\vec{j} }_p ( \vec{r} ) \rangle  \\
\nonumber
&& - \frac{e^2}{m} [ n ( \vec{r} ) {\vec{A} }_c ( \vec{r} ) + \langle
\hat{n} ( \vec{r} ) \vec{A} ( \vec{r} ) \rangle ] + \bm{ \nabla } \times
\langle \vec{M} ( \vec{r} ) \rangle . 
\end{eqnarray}
We notice that the last term in square brackets represents an unconventional
contribution  to the {\it classical} current, that might be observable in cases
of matter-field entanglement.   \par

The Schr\"odinger equation for the total system can be derived from the
functional $ S_Q $ of eq.(7), by variation with respect to $ \langle{ \Psi
}_T  (t) \vert $ . Using eq.s (11), (12) and dropping a global
time-dependent phase-factor, one obtains

\begin{equation}
i \hbar {\partial }_t \vert {\Psi }_T \rangle = ( H_0 + H_{int} ) \vert {\Psi
}_T \rangle ,
\end{equation}
where $ H_0 $, given by

\begin{eqnarray}
&& H_0 = H_M + H_F   \\ \nonumber
&& - \int d \vec{r} [ {\vec{A} }_c \cdot {\vec{j} }_p - \frac{e^2 }{2 m}
\hat{n} \vert {\vec{A} }_c {\vert }^2 + {\vec{B} }_c \cdot \vec{M} 
] ,
\end{eqnarray}
will be assumed as unperturbed Hamiltonian. We remark that the self-interaction
due to the coherent field is included in the unperturbed Hamiltonian $ H_0 $.
Assuming an unperturbed state of the form

\begin{equation}
\vert {\Psi }_T {\rangle }_0 = \vert \psi \rangle \otimes \vert 0 {\rangle }_F
,
\end{equation}
where $ \vert 0 {\rangle }_F $ is the vacuum state for the field, we can write
the unperturbed Schr\"odinger equation in the form

\begin{equation}
i \hbar {\partial }_t \psi = ( 2m )^{-1} \vert - i \hbar \bm{ \nabla } + e
{\vec{A} }_c {\vert }^2 \psi + {\mu }_B {\vec{B} }_c \cdot {\bm{\sigma} } \:
\psi .
\end{equation}
The non-linearity of of the Schr\"odinger equation (17) is evident if the
eqs. (11) and (12) are taken into account. \par

The perturbative Hamiltonian $ H_{int} $ of eq.(14) is given by

\begin{eqnarray}
&& H_{int} = - \int d \vec{r} \vec{A} \cdot [ ( {\vec{j} }_p - \langle 
{\vec{j} }_p \rangle ) - \frac{e^2 }{m} ( \hat{n} - \langle \hat{n} \rangle )
{\vec{A} }_c  \\ \nonumber
&& + \frac{e^2 }{m} \langle \hat{n} \vec{A} \rangle ]   + \frac{e^2 }{2 m}
\int d \vec{r} \hat{n} \vert {\vec{A} }_c {\vert }^2 - \int d \vec{r}
\vec{B} \cdot ( \vec{M}  - \langle \vec{M}  \rangle ) .
\end{eqnarray}

The Hamiltonian $ H_{int} $ contains the residual uncoherent interaction with
the quantum field. It describes processes like uncoherent  real photon emission
as well as virtual photon emission and reabsorption (Lamb shift). Such
processes will not be analysed in this paper. A justification for this  will be
given at the end of Sec.V. Furthermore, for the sake of simplicity, in what
follows the spin interaction will be neglected.   \par

Finally we observe that the Schr\"odinger equation (17) and the Maxwell
equation (12) can be derived from the following Lagrangian density

\begin{eqnarray} 
{\cal L} &&= i \hbar {\psi }^* {\partial }_t \psi   - (2m)^{-1} \vert ( - i
\hbar \bm{\nabla } + e {\vec{A} }_c ) \psi {\vert }^2 \\ \nonumber 
&&+ ( {\epsilon }_0 /2 ) ( \vert \partial {\vec{A} }_c / {\partial t} {\vert
}^2 - c^2 \vert {\vec{B} }_c {\vert }^2 ) ,
\end{eqnarray}
where the Coulomb gauge is understood and the spin term has been neglected, for
the sake of simplicity.

\section{Conserved energy.}
\label{sec:3}

By the N\"other theorem we can obtain, from the Lagrangian density (19), 
expressions for the conserved mean energy of the system

\begin{eqnarray} 
{\cal E} &&= \int d \vec{r} {\psi }^* ( - \frac{ {\hbar }^2 }{2m}  {\nabla }^2
- \frac{i e \hbar }{m} {\vec{A} }_c \cdot \bm{\nabla } + \frac{ e^2 }{2m}
A_c^2 ) \psi  \\ \nonumber 
&&+ \frac{{\epsilon }_0 }{2} \int d \vec{r} ( E^2_{\perp c} + c^2 B^2_c ) 
\end{eqnarray}
as well as for the total momentum

\begin{equation}
\vec{P} = - i \hbar \int d \vec{r} {\psi }^* \bm{\nabla } \psi + {\epsilon
}_0 {\sum }_j \int d \vec{r} E_{\perp c j } \bm{\nabla } A_{c j} .
\end{equation}

For a localized wave-packet the conservation of $ {\cal E } $ can be verified
directly. In fact, by use of eq.(17), we obtain

\begin{eqnarray}
&& \frac{ d {\cal E} }{d t} = \int d \vec{r} \frac{\partial {\vec{A} }_c
}{\partial t} \cdot \{ - \frac{i e \hbar }{2m} ( {\psi }^* \bm{\nabla } \psi
- \psi \bm{\nabla } {\psi }^* )      \\ \nonumber 
&& + \frac{ e^2 }{m} {\vec{A} }_c \vert \psi 
{\vert }^2 - {\epsilon }_0 \frac{\partial {\vec{E} }_{\perp c} }{\partial t} +
{\epsilon }_0 c^2 \bm{\nabla } \times {\vec{B} }_c  \\ \nonumber
&& - {\epsilon }_0 c^2 \bm{ \nabla } \cdot ( {\vec{E } }_c \times {\vec{B}
}_c ) \}  \\ \nonumber 
&& = - {\epsilon }_0 c^2 \oint d \vec{s} \cdot ( {\vec{E} }_{\perp c } \times
{\vec{B} }_c )  ,
\end{eqnarray}
where the r.h.s. represents the ingoing flux, at infinite distance, of the
Pojnting vector. In the absence of an appreciable emission of coherent
radiation, one obtains $ d{\cal E} / dt \simeq 0 $. In eq.(22) use has been
made of eq.s (11) and (12), as well as of the transversality of $ {\vec{A}
}_c $.  \par 
   
The last term in eq.(20), which represents the energy of the {\it classical}
field, can be cast in the form:

\begin{eqnarray} 
&& {\cal E}_F = \frac{ {\epsilon }_0 }{2} \int d \vec{r} ( E^2_{\perp c} + c^2
B^2_c )  \\ \nonumber  
&&= \int d \vec{r} {\psi }^* ( \frac{ i e \hbar }{2m} {\vec{A} }_c \cdot
\bm{\nabla } + \frac{ e^2 }{2m} A^2_c ) \psi + {\epsilon }_0 \int d \vec{r}
E^2_{\perp c} \\ \nonumber 
&& - \frac{ {\epsilon }_0 }{4} \frac{ d^2 }{d t^2 } \int d \vec{r} A^2_c +
\frac{1}{2} {\epsilon }_0 c^2 \oint d \vec{s}  \cdot ( {\vec{A} }_c \times
{\vec{B} }_c ). 
\end{eqnarray}
In the absence of a significant emission of coherent radiation, the last term
can be neglected for a finite system.  \par

By use of this result, we obtain from eq.(20)

\begin{eqnarray}
{\cal E} &&= - \int d \vec{r} {\psi }^* \frac{ {\hbar }^2 }{2m} {\nabla }^2
\psi - \frac{1}{2} \int d \vec{r} {\vec{A} }_c \cdot {\psi }^* \frac{ i e
\hbar }{m}  \bm{\nabla }  \psi \\ \nonumber 
&&+ {\epsilon }_0 \int d \vec{r} E^2_{\perp c} + \frac{ {\epsilon }_0 }{4}
\frac{d^2 }{d t^2} \int d \vec{r} A^2_c .
\end{eqnarray} 
Two features of eq.(24) are remarkable. The first one is the absence of $
A^2_c $ from the interaction term. The second one is the factor of $ 1/2 $
appearing in the interaction term (a proper result for a self-interaction
energy). \par

\section{Self-interaction energy for a gaussian wave-packet.}
\label{sec:4}

In order to simplify the mathematical analysis of the problem, we assume a wave
function of the form (a sort of de Broglie's double solution)

\begin{eqnarray}
\psi ( t, \vec{r} ) &&= \exp \{ i {\hbar }^{-1} [ \: {\vec{p} }_c (t) \cdot
\vec{r}  \\ \nonumber  &&- {\int }_0^t d t' \frac{ p^2_c ( t' ) }{2m} \:] \}
\; \phi (t, \vec{r} - {\vec{r} }_c ( t )  ) .
\end{eqnarray}
In eq.(25) the function $ \phi $ is chosen in such a way that the following
relations 

\begin{equation}
{\vec{p} }_c = - i \hbar \langle \psi \vert \bm{\nabla } \vert \psi \rangle
\end{equation}
and

\begin{equation}
{\vec{r} }_c = \langle \psi \vert \vec{r} \vert \psi \rangle
\end{equation}
hold. With these assumptions $ {\vec{p} }_c $ and $ {\vec{r} }_c $ can be
interpreted as {\it classical} momentum and position of the particle,
respectively. In order to simplify the calculations, we refer to a Gaussian
wave function

\begin{equation}
\phi ( \vec{r} ) = ( \frac{1}{4 \pi b^2 } {)}^{3/4} \: \exp [- \frac{ r^2 }{8
b^2 } ] ,
\end{equation}
as a model. \par

The time evolution for both $ {\vec{p} }_c $ and $ {\vec{r} }_c $ can be
obtained by the Ehrenfest theorem \cite{8} (for $ {\vec{r} }_c $ see eq.(35)
in the sequel, while $ {\vec{p} }_c $ is approximately conserved, according to
eq.(44) ). \par

The probability density $ \rho $, corresponding to the wave function (25), is
given by

\begin{equation} 
\rho ( t, \vec{r} ) \equiv {\rho }^0 ( t ,\vec{r} - {\vec{r} }_c
)  = \vert \phi ( t ,\vec{r} - {\vec{r} }_c ) {\vert }^2 ,
\end{equation}
or, in Fouri\'er representation, by

\begin{equation}
\hat{\rho } (t, \vec{q} ) = e^{-i \vec{q} \cdot {\vec{r} }_c } {\hat{\rho }
}^0 ( t, \vec{q} ) = e^{-i \vec{q} \cdot {\vec{r} }_c } e^{- b^2 q^2 } ,
\end{equation}
where the last expression refers to the Gaussian wave function of eq.(28).
\par

The solution to eq.(12) is given by the retarded potential

\begin{eqnarray}
&& {\vec{A} }_c ( t, \vec{r} ) = \frac{1}{4 \pi {\epsilon }_0 c^2 } \int d
{\vec{r} }' \frac{ {\vec{j} }_{c \perp } ( t - \tau , {\vec{r} }' ) }
{ \vert \vec{r} - {\vec{r} }' \vert }  \\ \nonumber
&& \simeq \frac{1}{4 \pi {\epsilon }_0 c^2 } \int d
{\vec{r} }' \frac{ {\vec{j} }_{c \perp } ( t, {\vec{r} }' ) }
{ \vert \vec{r} - {\vec{r} }' \vert }  ,
\end{eqnarray}
where we have assumed a density $ {\rho }_0 $ sufficiently localized in space,
in order that the effects depending on the retardation time

\begin{equation}
\tau = c^{-1} \vert \vec{r} - {\vec{r} }' \vert
\end{equation}
be negligible. We recall that, in the classical limit, the first order
contribution in $ \tau $ (actually neglected) is responsible for the radiation
reaction. For the sake of simplicity, the term containing the potential will be
neglected in the expression  (13) for $ {\vec{j} }_c $. In this way we
obtain for the {\it classical} current $ {\vec{j} }_c $ the following
approximate expression

\begin{eqnarray}
&& {\vec{j} }_c \simeq \langle {\vec{j} }_p \rangle  \\ \nonumber
&& = - e m^{-1} {\vec{p} }_c {\rho }^0 - i e \hbar (2m)^{-1} ( {\phi }^*
\bm{ \nabla } \phi - \phi \bm{\nabla } {\phi }^* ) \\ \nonumber
&& \simeq - e m^{-1} {\vec{p} }_c \, {\rho }^0 ( t, \vec{r} - {\vec{r} }_c
).
\end{eqnarray} 
In the last equality we have assumed a {\it classical} momentum $ {\vec{p} }_c
$ sufficiently large, in order that the current due to the {\it internal}
motion be negligible, compared with the {\it convective} one. \par

Let us define $ {\vec{A} }^0_c ( \vec{r} ) = {\vec{A} }_c ( \vec{r} +
{\vec{r} }_c ) $. From eq.s (30), (31) and (33) we obtain

\begin{equation}  
{\hat{{\vec{A} }}}^0_c \simeq - \frac{e}{m c^2 {\epsilon }_0 q^2 } {\hat{\rho
}}^0 ( \vec{q} ) [ {\vec{p} }_c - q^{-2} \vec{q} ( \vec{q} \cdot {\vec{p}
}_c ) ] .
\end{equation}
According to the Ehrenfest theorem, the {\it classical} velocity of the
electron is given by

\begin{equation}
{\vec{v} }_c \equiv d {\vec{r} }_c / d t = m^{-1} ( {\vec{p} }_c + e \langle
{\vec{A} }^0_c \rangle ) .
\end{equation}
where the average is taken over the {\it internal} wave function $ \phi $. We
obtain

\begin{eqnarray}
&&e \langle {\vec{A} }^0_c \rangle = e \int d \vec{r} {\rho }^0 ( \vec{r} ) 
{\vec{A} }^0_c ( \vec{r} )  \\ \nonumber 
&&= e ( 2 \pi )^{-3} \int d \vec{q} {\hat{\rho }}^{0 *} ( \vec{q} )
{\hat{\vec{A} }}^0_c ( \vec{q} ) \\ \nonumber 
&&= - \frac{e^2 }{m c^2 {\epsilon }_0 } (2 \pi )^{-3} \int d \vec{q} q^{-2}
\vert {\hat{\rho }}^0 ( \vec{q} ) {\vert }^2 [{\vec{p} }_c - q^{-2} \vec{q}
( \vec{q} \cdot {\vec{p} }_c ) ]  \\ \nonumber 
&&= - \frac{ e^2 {\vec{p} }_c }{m c^2 {\epsilon }_0 } \frac{8 \pi }{3} ( 2 \pi
)^{-3} \int d q \vert {\hat{\rho }}^0 ( \vec{q} ) {\vert }^2 \\ \nonumber 
&&= - \frac{4}{3} \frac{{\cal E}_{el} }{m c^2 } {\vec{p} }_c ,
\end{eqnarray}
where the {\it electrostatic} energy

\begin{eqnarray} {\cal E}_{el} &&= \frac{e^2}{8 \pi {\epsilon }_0 } (2 \pi
)^{-3} \int d \vec{q} 4 \pi q^{-2} \vert {\hat{\rho} }^0 ( \vec{q} ) {\vert
}^2 \\ \nonumber 
&&= \frac{e^2}{4 {\pi }^2 {\epsilon }_0 } \int d q \vert {\hat{\rho }}^0 (
\vec{q} ) {\vert }^2 = \frac{e^2}{8 \sqrt{2} {\pi }^{3/2} {\epsilon }_0 b }
\end{eqnarray}
has been introduced. The last expression refers to the Gaussian model. We
obtain in this way

\begin{equation}
{\vec{p} }_c \simeq m ( 1 + \frac{4}{3} \frac{{\cal E}_{el} }{ m c^2 } )
{\vec{v} }_c  .
\end{equation}
Eq.(38) shows that, as a consequence of the self-interaction, a mass
renormalization takes place. Exactly the same result holds for an extended
classical particle (notice the famous factor of $ 4/3 $ in the r.h.s. of
eq.(38)). \par

The {\it classical} transverse electric field is given by

\begin{eqnarray} 
&& {\hat{\vec{E} }}_{\perp c} (t,\vec{q} ) = - {\partial }_t {\hat{\vec{A}
}}_c (t, \vec{q} )  \\ \nonumber 
&&= \frac{e \, e^{-i \vec{q} \cdot {\vec{r} }_c } }{{\epsilon }_0 m c^2 q^2 }
{\hat{\rho }}^0 ( t, \vec{q} ) [ {\vec{p} }_c - q^{-2} \vec{q} ( \vec{q}
\cdot {\vec{p} }_c ) ] [ i \vec{q} \cdot {\vec{v} }_c  - {\partial }_t \ln
{\hat{\rho }}^0 ].  \\ \nonumber
&& \simeq i \frac{e \, e^{-i \vec{q} \cdot {\vec{r} }_c } }{{\epsilon }_0 m
c^2 q^2 } {\hat{\rho }}^0 ( t, \vec{q} ) [ {\vec{p} }_c - q^{-2} \vec{q} (
\vec{q} \cdot {\vec{p} }_c ) ]  \vec{q} \cdot {\vec{v} }_c  ,
\end{eqnarray}
where the contribution due to the time variation of $ \ln {\hat{\rho } }^0 $
has been neglected, with respect to the contribution due to the convective
motion.  \par

In what follows the r.h.s. of eq.(24) will be calculated up to the second
order in the electron charge $ e $. Moreover any power of $ \beta = v_c /c $ 
higher than the second will be neglected with respect to unity. \par

Let us proceed to calculate the r.h.s. of eq.(24). The first term is given by

\begin{eqnarray}
&& - \frac{{\hbar }^2 }{2 m} \int d \vec{r} {\psi }^* {\nabla }^2 \psi  \\
\nonumber 
&&= \frac{ p^2_c }{2m} - \frac{{\hbar }^2 }{2m} \int d \vec{r} {\phi }^*
{\nabla }^2 \phi = \frac{ p^2_c }{2m} + \frac{3 {\hbar }^2}{ 16 m b^2 } .
\end{eqnarray}
The next term is given by

\begin{eqnarray}
&& - \frac{1}{2} \int d \vec{r} \, {\vec{j} }_c \cdot {\vec{A} }_c  = -
\frac{1}{16 {\pi }^3 } \int d \vec{q} \, {\hat{\vec{j} } }^*_c  (
\vec{q} ) \cdot {\hat{\vec{A} } }_c ( \vec{q} )  \\ \nonumber
&& = - \frac{e^2 v^2_c }{16 {\pi }^3 {\epsilon }_0 c^2 } \int d \vec{q} \,
q^{-2} \vert {\hat{\rho } }_0 ( q ) {\vert }^2 [ 1 - \frac{( \vec{q} \cdot
{\vec{p} }_c )^2 }{q^2 p^2_c } ] \simeq - \frac{2}{3} \: {\beta }^2 {\cal
E}_{el} .
\end{eqnarray}
Next let us calculate 

\begin{equation}
{\epsilon }_0 \int d \vec{r} E^2_{\perp c} \simeq  \frac{4}{15} {\beta }^4 
{\cal E}_{el}.
\end{equation}
This term, of the order of $ {\beta }^4 $, is negligible. The last term 
consists of

\begin{eqnarray} 
&& - \frac{{\epsilon }_0 }{4} \frac{d^2 }{d t^2 }\int d \vec{r} A^2_c  \simeq
\frac{ {\beta }^2 }{c^2 } \frac{d }{d t } ( \frac{4}{3} {\cal E}_{el} b \frac{d
b }{d t } )   \\ \nonumber
&& \simeq  \frac{8}{3} \sqrt{ \frac{2}{\pi } } \frac{ {\beta }^2 }{c^2 }{\cal
E}_{el} b \frac{d b }{d t } ) 
\end{eqnarray}
and is negligible also. \par

Finally let us calculate the total momentum $ \vec{P} $ of eq.(21), which is
conserved according to the N\"other theorem. We obtain

\begin{eqnarray}
\vec{P} && = {\vec{p} }_c + i {\epsilon }_0 ( 2 \pi )^{-3} {\sum }_j \int d
\vec{q} \vec{q} {\hat{E} }^*_{\perp c j } {\hat{A} }_{cj}( \vec{q} )  \\
\nonumber  
&&\simeq {\vec{p} }_c [ 1 + \frac{4}{15} {\beta }^2 \frac{ {\cal E}_{el}
}{mc^2} ] .
\end{eqnarray} 
Using this result we obtain for the first term of eq.(40)

\begin{eqnarray}
\frac{p^2_c }{2m} && \simeq \frac{P^2 }{2m} [ 1 - \frac{8}{15} {\beta }^2
\frac{{\cal E}_{el} }{m c^2 } ]  \\ \nonumber  
&&\simeq \frac{P^2 }{2m} - \frac{4}{15} {\beta }^4 {\cal E}_{el} \simeq
\frac{P^2}{2m} = const.
\end{eqnarray}
This means that the first term in the r.h.s. of eq.(40) is approximately a
constant of the motion. \par

Finally the total energy of eq.(24) is given by the following expression

\begin{equation}
{\cal E} \simeq  \frac{1}{2} m v^2_c - \frac{ {\hbar }^2 }{2m} \int d \vec{r}
{\phi }^* {\nabla }^2 \phi - \frac{4}{3} \beta {\cal E}_{el} .
\end{equation}

\section{Localization and binding energy}
\label{sec:5}

From eq.(46) we obtain, for the Gaussian wave function of eq.(28),

\begin{equation} 
{\cal E} \simeq \frac{1}{2} m v^2_c + \frac{3 {\hbar }^2 }{16 m b^2 }
-\frac{e^2 {\beta }^2 }{6 \sqrt{2 \pi } {\epsilon }_0 b } .
\end{equation}
A minimum for $ {\cal E} $ is attained for $ b $ given by

\begin{equation} 
b_{el} \simeq \frac{9 {\pi }^{3/2 } {\epsilon }_0 {\hbar }^2 {\beta }^{-2 } }{
\sqrt{2} m e^2 } = \frac{9 \sqrt{\pi } }{4 \sqrt{2 } } {\beta }^{-2} \, a_B
\simeq 2.8 {\beta }^{-2} \, a_B ,
\end{equation}
where $ a_B = 4 \pi {\epsilon }_0 {\hbar }^2 / m e^2 $ is the Bohr radius. The
corresponding {\it binding} energy is given by

\begin{equation}
{\cal E}^b_{el} \simeq (4 / 27 \pi ) {\beta }^4 E_R ,
\end{equation}
where $ E_R = m e^4 / 2 ( 4 \pi {\epsilon }_0 \hbar )^2 $ is the Rydberg
energy. Indicatively, for $ \beta \sim 10^{-1} $ we obtain $ b_{el} \sim 1.5
\times 10^{-8}$ m and $ {\cal E}^b_{el} \sim  6.4 \times 10^{-5} $ eV. \par

For a particle with charge $ \pm Ze $ and mass $ M $ eq.s (48) and (49) read

\begin{equation}
b \simeq (m /M ) Z^{-2} b_{el}
\end{equation}
and 

\begin{equation}
{\cal E}^b \simeq (M / m) Z^4 {\cal E}^b_{el} .
\end{equation}
Indicatively, for a proton with $ \beta \sim 10^{-1} $ one obtains $ b \sim 8.1
\times 10^{-12} $ m and $ {\cal E}^b \sim 1.2 \times 10^{-1} $ eV. \par

We observe that, for fixed $ \beta $, the ratio between the localization radius
$ b $ and the de Broglie wave-length $ \lambda $ is independent of the mass,
according to 

\begin{equation}
b / {\lambda } \simeq ( 2.8 \; a_B \, / \, 2 \pi {\lambda }_c ) \; {\beta
}^{-1} Z^{-2} \simeq 62 \; {\beta }^{-1} Z^{-2} ,
\end{equation}
where $ {\lambda }_c $ is the Compton wave-length for the electron.  \par
 
All of these results have been obtained by assuming an isotropic Gaussian wave
function  $ \phi $. It can be expected that the lowest-energy configuration
would not correspond to a spherically symmetric wave function $ \phi $, but
rather  to a cylindrically symmetric one.  \par

We recall that the interaction of the system with the residual quantum-field,
described by $ H_{int} $ of eq.(18), has been neglected in this paper. This
amounts to neglect the emission of real uncoherent photons by the system, as
well as the emission and reabsorption of virtual photons. In both processes the
field is coupled essentially to the {\it internal} motion and not to the {\it
convective} one, as shown by eq.s (18) and (33). We remark that the emission of
uncoherent radiation is a dissipative effect. Presumably it plays a r\^ole in
the relaxation of the system toward a bound state, but it cannot increase the
internal energy or destroy the bound state itself. On the other hand, the
emission and reabsorption of virtual photons can be expected to introduce a
very small correction to the unperturbed binding energy, of the order of the
Lamb shift energy for a system, whose space dimension is of the order of $ b $
given by eq.(48) or by eq.(50).

\section{Neutral atom.}
\label{sec:6}

The charge density for a neutral atom is given by

\begin{equation}
{\rho }_{ch} (\vec{r} ) = Z e {\rho }_{cm} (\vec{r} ) - Z e \int d { \vec{r}
}' {\rho }_{cm} ( {\vec{r} }' ) {\rho }_{el} ( \vec{r} - {\vec{r} }' ) ,
\end{equation}
where $ {\rho }_{cm} $ represents the probability density for the
centre-of-mass co-ordinate (coinciding approximately with the nuclear
co-ordinate) and $ Z {\rho }_{el} $ is the electron density referred to the
nuclear position. In Fouri\'er representation eq.(53) reads

\begin{equation}
{\hat{\rho } }_{ch} ( \vec{q} ) = Z e {\hat{\rho} }_{cm} ( \vec{q} ) \, [ 1 -
{\hat{\rho } }_{el} ( \vec{q} ) ] .
\end{equation}
If, for the sake of simplicity, we assume a Gaussian form for both $ {\hat{\rho
} }_{cm} $ and $ {\hat{\rho } }_{el} $, the {\it electrostatic} energy $ {\cal
E}_{el} $ is given by

\begin{eqnarray}  
{\cal E}_{el} &&\simeq \frac{ Z^2 e^2 }{4 {\pi }^2 {\epsilon }_0 } {\int
}_0^{\infty } d q \vert {\hat{\rho} }_{cm} {\vert }^2 \, \vert 1 - {\hat{\rho }
}_{el} {\vert }^2  \\ \nonumber 
&&= \frac{ Z^2 e^2 }{4 {\pi }^2 {\epsilon }_0 } {\int }_0^{\infty } d q e^{ -2
b^2 q^2 } \, [ 1 - e^{- {\gamma }^2 q^2 } ]^2 \\ \nonumber 
&&= \frac{ Z^2 e^2 }{8 \sqrt{2} {\pi }^{3/2 } {\epsilon }_0 } \, [ \frac{1}{b}
- \frac{ 2 \sqrt{2} }{ \sqrt{2 b^2 + {\gamma }^2 } }  + \frac{1}{ \sqrt{ b^2 +
{\gamma }^2 } } ] .
\end{eqnarray} 
For a delocalized centre-of-mass, {\it i.e.} for $ b \gg \gamma $, one obtains
$ {\cal E}_{el} \simeq 0 $, as for a neutral particle. On the contrary, for a
strong localization of the centre-of-mass, {\it i.e.} for $ b \ll \gamma $, one
obtains

\begin{equation}
{\cal E} \simeq \frac{ Z^2 e^2 }{8 \sqrt{2} {\pi }^{ 3/2} {\epsilon }_0 b } ,
\end{equation}
as for a bare nucleus. In this case eq.s (50) and (51) as well as (52) hold.

\section{Conclusion}
\label{sec:7}

We have shown the existence of a mechanism which favours a spontaneous
localization in space for a moving charged particle. The effect is due
essentially to the attractive force acting among parallel currents. It is well
known that in quantum mechanics this force, for a single particle, is not
contrasted by a corresponding repulsion due to the Coulomb field (as it
happens, on the contrary, for an extended classical particle). In fact, the
interaction of the electromagnetic field with non-relativistic matter is
described in a simple way assuming the Coulomb gauge. In such a description,
the Coulomb interaction is viewed as a direct instantaneous action among
different particles, as {\it e.g.} in the hydrogen atom. No electrostatic
self-interaction is assumed for a single particle. It is known, from the very
beginning of wave mechanics, that such a kind of interaction, if present, would
shift the energy eigenvalues of the hydrogen atom to physically wrong values
\cite{9}. Furthermore one can convince himself easily that, if absurdely such a
kind of self-interaction should exist, the average energy of an atom would
depend on the localization of its centre-of-mass. As a consequence of this fact
{\it e.g.} the Van der Waals crystals could not exist. \par

Some comments are due about the symmetry-breaking processes involved in the
localization effect. First of all a break-down of the translational invariance
is involved. Physically it can be explained by the following argument. A
physical electron, in a given {\it state} of momentum $ \vec{p} $ and helicity
$ s $, is represented by the minimal-energy {\it state} belonging to the sector
corresponding to charge $ -e $, total  momentum $ \vec{p} $ and  helicity $ s $
of the Hilbert space describing the total system (consisting of {\it bare}
electron plus field).From this point of view it is obvious that, as a
consequence of the interaction, the momentum must be shared between the field
and the {\it bare} electron. One can say in other words that the momentum
density does not coincide with the charge density, since part of the momentum
is due to the field, which is neutral.  This implies a localization for the
charge.    \par

From the point of view of special relativity, a more intiguing consequence
stems  from the velocity dependence of the localization effect. In fact a
non-relativistic model, like the one analysed in this paper, should be imagined
as a low-energy limit of some hypothetical Lorentz invariant one. From such a
point of view it is evident that, as the spontaneous localization effect brakes
down the translational invariance, so the velocity dependence of the effect can
be interpreted as an indication for a spontaneous breakdown of the Lorentz
invariance. On the other hand it is a known result, in relativistic QED, that
the Lorentz invariance is broken in any charged sector of the Hilbert space
(see {\it e.g.} Ref.\cite{10}). In this context the result obtained in this
paper appairs as a confirmation, in a non-relativistic situation, of the
theorem quoted above. The question arises if this result represents a 
mathematical strangeness only, or if it can lead really to possible physical
observations conflicting with special relativity. A possible loophole to save
the Lorentz invariance may be the following one. It may be possible that
repeated preparations of electrons would in practice prepare different states,
in which (partially) localized electrons are randomly distributed over the
space region which supports the usual (non-localized) wave-function (a similar
situation is assumed in Bohmian mechanics). In such a case the Lorentz
invariance would not be violated by direct physical observations. However we
remark that simple energetical considerations, like the ones  developed in this
paper, are inadequate to clarify this point completely.   \par

Nevertheless, effects due to the localization, or more generally to the
interaction with the self-generated coherent field, could be observed
indirectly, through dynamical effects like {\it e.g.} spin dynamics, or more
directly in condensed matter, where a preferred reference frame exists
\cite{11}.

\end{document}